\documentclass[]{an}
\usepackage{graphicx}
\usepackage{times}
\usepackage{rotating}
\Volume{322}             % volume number
\Year{2001}              % format: {YYYY}
\Month{10}               % format: {M} or {MM} (number)
\Pagespan{223}{227}      % format: {start page}{last page}

\begin{document}
\lhead[\thepage]{A.N. Author: Title}
\rhead[Astron. Nachr./AN~{\bf XXX} (200X) X]{\thepage}
\headnote{Astron. Nachr./AN {\bf 32X} (200X) X, XXX--XXX}

\title{Optical and infrared photometry of new very low-mass stars and brown dwarfs in the $\sigma$ Orionis cluster}

\author{V. J. S. B\'ejar\inst{1} 
\and  M. R. Zapatero Osorio\inst{2}
\and  R. Rebolo\inst{1,3}
}
\institute{
Instituto de Astrof\'{\i}sica de Canarias, E-38205 La Laguna, Tenerife, Spain
\and LAEFF-INTA, P.O. Box 50727, E-28080 Madrid, Spain
\and Consejo Superior de Investigaciones Cient\'{\i}ficas, Madrid, Spain}

\date{Received {\it date will be inserted by the editor}; 
accepted {\it date will be inserted by the editor}} 

\abstract{We present an $RI$ photometric survey 
covering an area of 430 arcmin$^2$ around the multiple star 
$\sigma$ Orionis. The observations were conducted with the 0.8 m IAC-80 Telescope at the Teide Observatory. The survey limiting $R$ and $I$ magnitudes are 22.5 and 21, and completeness 
magnitudes 21 and 20, respectively. We have selected 53 candidates from the $I$ vs.\ $R$--$I$ 
colour-magnitude diagram ($I$ = 14--20) that follow the previously known photometric sequence of the cluster. 
Adopting an age of 2--4 Myr for the cluster, we find that these objects span a mass range from 0.35 $M_{\odot}$ to 0.015 $M_{\odot}$. 
We have performed $J$-band photometry of 52 candidates and $K_{\rm s}$ photometry for 12 of them, with the
result that 50 follow the expected infrared sequence for the cluster, thus 
confirming with great confidence that the majority of the candidates are bona fide members. $JHK_{\rm s}$ photometry 
from the Two Micron All Sky Survey (2MASS) is available for 50 of the candidates and are 
in good agreement with our data. Out of 48 candidates, which have photometric accuracies better than
0.1 mag in all bands, only three appear to show near-infrared excesses.
}

\correspondence{vbejar@ll.iac.es}

\maketitle

\section{Introduction}

The Orion complex, because of its relative closeness and youth, is one of the most suitable sites for understanding low-mass star 
formation processes.
Recently, {\itshape ROSAT} pointed observations within this complex has led to the discovery of the 
very young stellar cluster $\sigma$ Orionis, around the multiple star of the same name (Walter et al. 
\cite{walter97}; Wolk \& Walter \cite{wolk98}).
Follow-up photometric and spectroscopic studies have revealed a sequence of objects in the colour--magnitude diagram
that extends well below the substellar limit (B\'ejar et al. \cite{bejar99}, hereafter BZOR; B\'ejar et al. \cite{bejar01}, hereafter BMZO). 
Studies of the depletion of lithium in the atmosphere of K6--M8.5 type low-mass members of the cluster 
impose an upper limit of 8 Myr on the age and suggest a most likely cluster age  
in the interval 2--4 Myr (Zapatero Osorio et al. \cite{osorio02}). 
{\itshape Hipparcos} provides a distance modulus of $m-M = 7.7 \pm 0.7$ (Perryman et al. \cite{perryman97}) 
for the central star, $\sigma$ Orionis. 
This star is affected by a low extinction value of $E(B-V) = 0.05$ (Lee \cite{lee68}), and 
the associated cluster also seems to exhibit very little reddening (see  
BMZO and Oliveira et al. \cite{oliveira02}). 
These combined characteristics of youth, proximity and low extinction make $\sigma$ Orionis one of 
the most interesting clusters for studying young substellar objects and the substellar mass function.

In this paper we present an extension of the $RI$ survey conducted by  BZOR, 
with the aim of detecting new low-mass star and brown dwarf candidates in 
the $\sigma$ Orionis cluster. 
We also present near-infrared data for these objects.  
Details of the observations are 
indicated in Section 2. In sections 3.1 and 3.2 
%and 3.3 
we explain the selection of the candidates and discuss 
their membership 
%, spatial distribution in the cluster 
and the presence of near-infrared excesses. Our conclusions are given in section 4.

\section{Observations}

\subsection{Optical Photometry}

\begin{figure}
\resizebox{\hsize}{!}
{\includegraphics[]{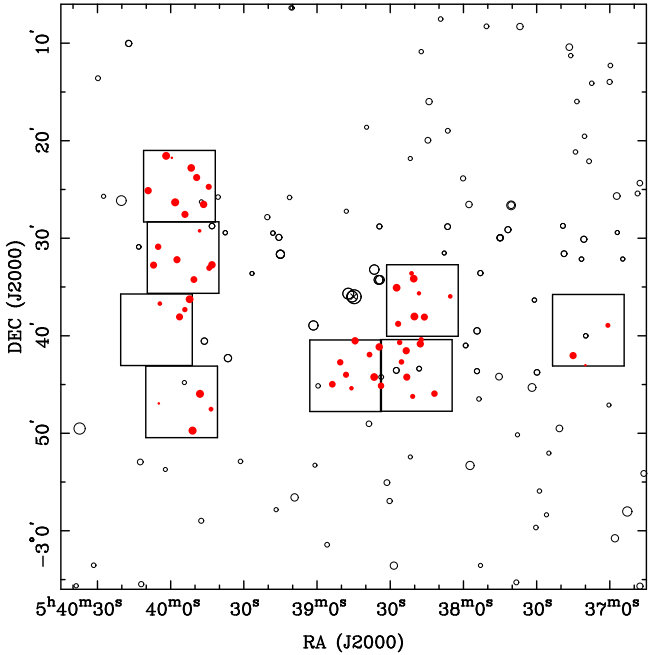}}
\caption{Location of surveyed fields (open squares) around the star $\sigma$ Orionis. Our candidates (from Table \ref{tab2}) are indicated 
by filled circles, while open circles denote field stars brighter than 13 mag. Relative brightness is 
indicated by symbol size. North is up and East is to the left.}
\label{fig1}
\end{figure}

We obtained $RI$ images with the IAC-80 Telescope, at the Teide Observatory on 1998 January 
22 and 23. The camera consists of a 1024 $\times$ 1024 Thomson CCD detector, 
providing a pixel projection of 0.4325 arcsec and a field of view of 
7.4 $\times$ 7.4 arcmin$^2$ in each exposure. We observed eight different fields in the two filters, 
covering a total area of 430 arcmin$^2$. Exposure times were 1800 s in each filter and field. 
Table \ref{tab1} lists the central coordinates of the eight fields. 
The location of the surveyed region is shown in Figure \ref{fig1}, where some 
bright stars are indicated. The star $\sigma$ Orionis, shown at the centre, is also included. 

\begin{table}[h]
\caption{Fields coordinates}
\label{tab1}
%\begin{center}
\begin{tabular}{ccc}\hline
Field  & R.A. (J2000) & Dec. (J2000)\\ 
      & (h m s)	 & ($^{\circ}$ $'$ $''$)	 \\ 
\hline
1&5 38 16.8&  $-$2 36 23\\ 
2&5 38 48.3&  $-$2 44 06\\ 
3&5 38 19.3&  $-$2 44 04\\ 
4&5 39 54.9&  $-$2 31 59\\ 
5&5 40 05.8&  $-$2 39 23\\
6&5 39 56.4&  $-$2 24 40\\
7&5 39 55.5&  $-$2 46 47\\ 
8&5 37 08.8&  $-$2 39 26\\
\hline
\end{tabular}
%\end{center}
\end{table}

Raw frames  were reduced within the {\sc IRAF}\footnote{IRAF is distributed by National Optical Astronomy
Observatories, which is operated by the Association of Universities for
Research in Astronomy, Inc., under contract with the National Science
Foundation.} environment, using the {\sc CCDRED} package.
Images were bias-subtracted and flatfield-corrected. We combined sky flats taken at dusk and dawn to obtain the flatfields. The photometric analysis was performed using routines within {\sc DAOPHOT}, 
including the selection of objects with a stellar point spread function (PSF) using the {\sc DAOFIND} task (extended objects were mostly avoided) 
and aperture and PSF photometry. The nights were photometric, and instrumental magnitudes were
transformed into the Cousins $RI$ system using observations of standard stars from Landolt (1992) made every night. Average seeing ranged from 1.5 to 2 arcsec. The survey completeness and limiting
magnitudes were $R$ = 21, $I$ =20 and $R$ = 22.5, $I$ = 21, respectively. 
We adopted as the completeness magnitude the value at which the histogram 
of detections as a function of magnitude reaches maximum ($\sim$ 10-$\sigma$ detection); 
and as limiting magnitude the value at which less than 50\%~of the objects at the maximum of the 
histogram are detected($\sim$ 3-$\sigma$ detection limit). 

\begin{figure}
\resizebox{\hsize}{!}
{\includegraphics[]{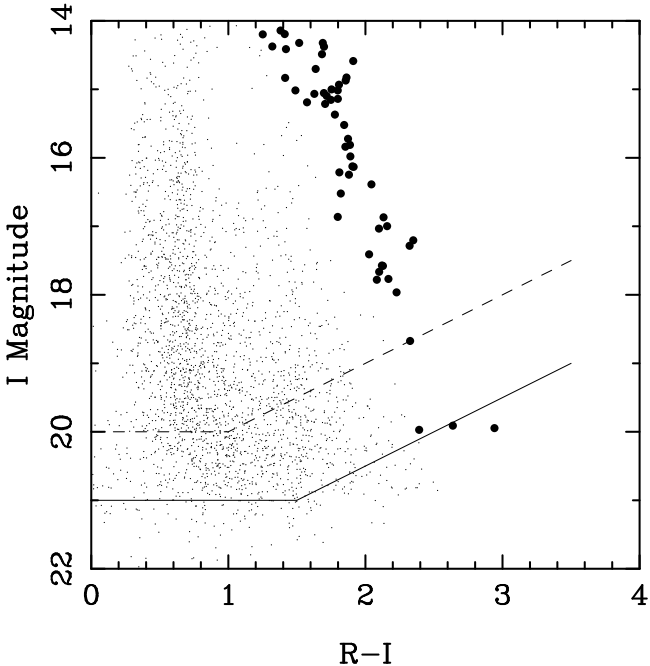}}
\caption{$I$ vs. $R-I$ colour--magnitude diagram in the $\sigma$ Orionis cluster 
resulting from our survey. Filled circles denote our selected candidates listed in Table \ref{tab2}. 
Completeness and limiting magnitudes are 
indicated by a dashed and solid line, respectively.}
\label{fig2}
\end{figure}

We have constructed $I$ vs. $R-I$ colour--magnitude  diagrams for each field in order to identify 
cool cluster members. Figure \ref{fig2} shows 
the combination of these diagrams for all the fields. We consider as candidate members objects redder and brighter 
than the field stars, following the cluster sequence previously defined in BZOR. 
 The lower envelope to the photometric sequence delineated by previously 
spectroscopically confirmed members is used to separate our candidates 
from interlopers and background sources. We have selected 53 candidate 
members, five of which are known from previous surveys (BZOR). Their magnitudes are in the range 
$I$ = 14--20 mag, which, according to recent theoretical models, correspond to masses in the range 0.35--0.015 $M_{\odot}$ 
(D'Antona \& Mazzitelli \cite{dantona97}; \cite{dantona98}; Burrows et al. \cite{burrows97}; 
Baraffe et al. \cite{baraffe98}; Chabrier et al. \cite{chabrier00}). The substellar mass limit at the age and distance of 
the $\sigma$ Orionis cluster is located at I $\sim$ 16\,mag. Table \ref{tab2} contains the list of selected 
candidates: around 31 are stellar and 22 substellar. 
Photometric data and coordinates 
are also included. Error bars account for the IRAF magnitude 
error and the uncertainty of the photometric calibration, which is typically $\pm$0.04-0.06 mag. Astrometry was carried out using the USNO-SA2.0 catalogue (Monet et al. \cite{monet96}); we estimate having 
achieved a precision better than 2 arcsec. Finder charts (3.7 $\times$ 3.7 arcmin$^2$) are provided 
in Figures \ref{figA1} and \ref{figA2}. 

\subsection{Infrared Photometry}

We obtained $J$- and $K_{\rm s}$-band point observations of the selected candidates 
with the  1.52 m Carlos S\'anchez Telescope (TCS), at the Teide Observatory, 
on 1998 September 18, December 17, 1999 January 23, 24 and 2000 January 27. 
The infrared camera (CAIN) is equipped with an 
HgCdTe 256 $\times$ 256 detector (Nicmos 3), which, with its wide optics configuration, provides a pixel projection of 1.00 arcsec, 
covering an area of 4.3 $\times$ 4.3 arcmin$^2$ in each exposure. Total exposure times ranged 
from 60 to 720 s depending on the filter and the expected magnitude of the candidates. 

Raw data were processed within the {\sc IRAF} environment. Each frame consisted of 9--10 exposures 
obtained using a dithering pattern on the detector. Final images were obtained 
combining individual images, properly aligned and sky-subtracted. Aperture photometry was 
performed for each object using the {\sc PHOT} routine within {\sc DAOPHOT}. A typical radial aperture of 
4--5 times the FWHM was adopted. Weather conditions were photometric. Average seeing ranged from 1.6 to 2.2 arcsec. In order to transform instrumental 
magnitudes into the UKIRT system, each night we observed several field standards (Hunt et al. \cite{hunt98}) and the Pleiades 
brown dwarf Calar3 (Zapatero Osorio, Mart\'{\i}n \& Rebolo \cite{osorio97a}).

\begin{figure}
\resizebox{\hsize}{!}
{\includegraphics[]{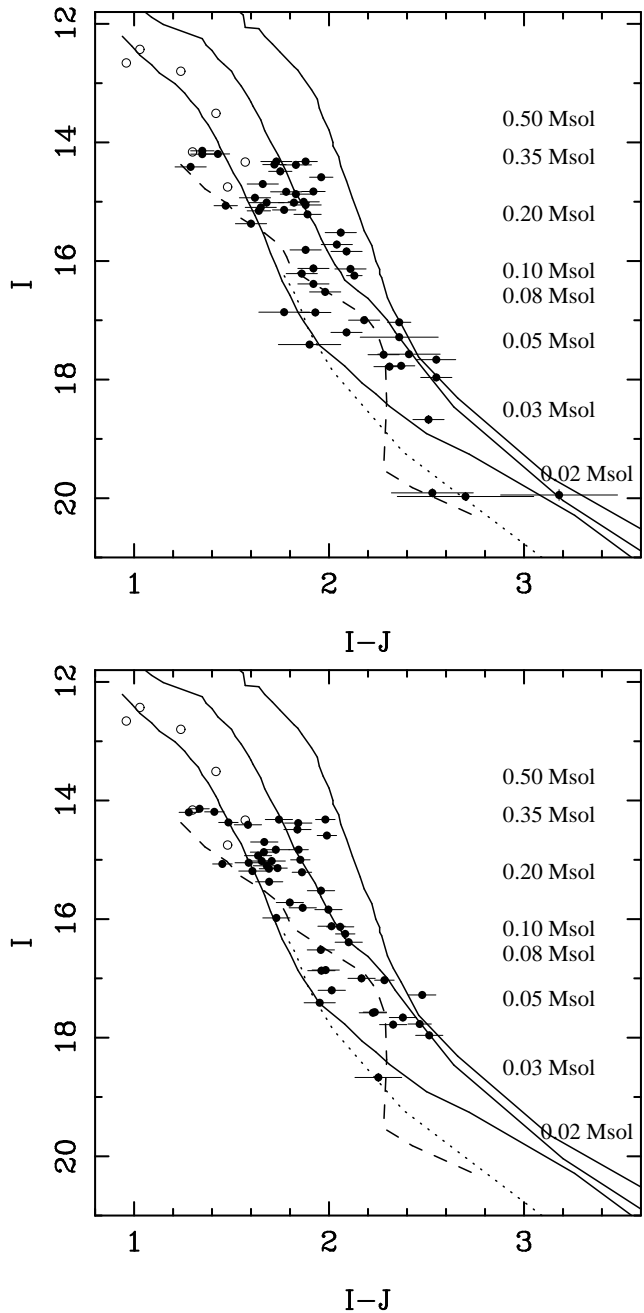}}
\caption{Top panel: $I$ vs. $I-J$ colour--magnitude diagram of selected candidates in our survey 
(filled symbols). Open circles denote previously known members taken from Wolk (\cite{wolk96}). 
For comparison, the 1, 3 and 10 Myr Next-Gen theoretical isochrones (solid lines from right to left), from the 
Lyon group (Baraffe et al. \cite{baraffe98}), new 3 Myr dusty isochrones (dotted line) from the 
Lyon group (Chabrier et al. \cite{chabrier00}) and models (dashed line) from D'Antona \& Mazzitelli (\cite{dantona98}) 
are also indicated. Bottom panel: $I$ vs. $I-J$ colour--magnitude diagram of selected candidates in our survey. As 
in previous figure, but $J$-band photometry taken from 2MASS.}
\label{fig3}
\end{figure}

\begin{figure}
\resizebox{\hsize}{!}
{\includegraphics[]{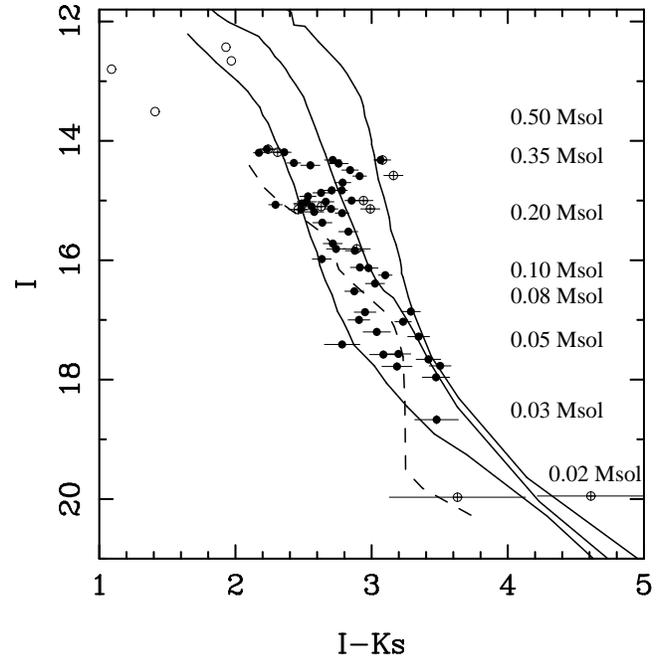}}
\caption{$I$ vs.\ $I-K_{s}$ colour--magnitude diagram of selected candidates. Filled symbols indicate those with 
2MASS photometry and open circles with error bars those with TCS photometry. Open circles denote previously 
known members taken from Wolk (\cite{wolk96}). 
For comparison, the same theoretical isochrones as in the previous figure are plotted.}
\label{fig4}
\end{figure}

The photometry of the candidates is shown in Table \ref{tab2}. Error bars include the IRAF magnitude 
error and the uncertainty of the photometric calibration, which is typically $\pm$0.05 mag. 
For 50 of the candidates, all except three, $JHK_{\rm s}$ photometry is available 
in the Two Micron All Sky Survey (2MASS) 
Point Source Catalogue, Third Incremental Release (Cutri et al. \cite{cutri03}). These data are also listed 
in Table \ref{tab2}. We estimate that 
the average difference between the 2MASS and TCS photometry in the $J$ band is 
 J(2MASS)-J(TCS)$=$0.012 $\pm$ 0.017 (where 0.017 stands for the error of the mean, the standard 
deviation of the differences is 0.116). For the ten candidates for which
we have $K_{\rm s}$-band photometry in common, the average difference is 
 $K_{\rm s}$(2MASS)-$K_{\rm s}$(TCS)$=$0.080 $\pm$ 0.040 (where 0.040 stands for the error of the mean, 
the standard deviation of the differences is  0.115). For eight of the objects the  
differences are larger than 0.2 mag in any of the two filters. One is an unresolved binary 
and other has large error bars in the 2MASS photometry. Small offsets between both data sets could be caused by differences 
in the filter systems, in the shape and strength of water absorption bands above the observatories, or due to the intrinsic variability in the atmospheres 
of some of the targets. In fact, one of the objects with a difference larger than 0.2 mag, 
S\,Ori J053825.4-024241, is known to present a strong photometric variability (Caballero et al. \cite{caballero04}). 
Figure \ref{fig3} shows the $I$ vs. $I-J$ diagrams for the selected candidates with 
our photometry (top panel) together with that obtained from the 2MASS 
catalogue (bottom panel). Figure \ref{fig4} shows a $I$ vs. $I-K_{\rm s}$ diagram for those objects with available $K_{\rm s}$ 
photometry.

\section{Discussion}
\subsection{Selection of Candidates and Cluster Membership}

Colour--magnitude diagrams based on the optical filters $R$, $I$ and $Z$ have 
proved to be a good technique for distinguishing true low-mass members from background 
objects in young nearby clusters (Prosser \cite{proser94}; Zapatero Osorio et al. \cite{osorio99}; BZOR ; 
Bouvier et al. \cite{bouvier98}). We have selected our cluster candidates using optical colour--magnitude diagrams 
in the $R$ and $I$ bands.  
The most important sources of contamination in our survey are field M dwarfs. Galaxies are 
mostly resolved within our completeness magnitude and, given the galactic latitude of the $\sigma$ 
Orionis cluster ($b=-17.3$ deg), giant stars are not expected to contribute in a significant number ($<$ 5\%) in comparison with
main-sequence dwarf stars (Kirkpatrick et al. \cite{kirketal94}). 
According to the density of M field dwarfs obtained by Kirkpatrick et al. (\cite{kirketal94}), we expect 
that our proposed photometric sequence for the cluster is not contaminated by field dwarfs of spectal type earlier than M4 
and by no more than three of later spectral type within the completeness of our survey. 
Contamination becomes important in the fainter part of 
our diagram, where, in addition to the larger error bars, foreground objects can be located in the 
cluster sequence reddened by the interstellar medium. 

To discriminate between bona fide cluster members and field objects, either spectroscopic 
data or infrared photometry is required. The advantage of the latter is that it can be 
performed in a shorter integration time in relative small telescopes. 
The combination of optical and infrared data is a trustworthy  technique for distinguishing bona 
fide cool cluster members (Zapatero Osorio et al. \cite{osorio97a}; Mart\'{\i}n et al. \cite{martin00}; BMZO). 
The membership of most of the low-mass stars and brown dwarfs ($>$ 90\%)  
identified using both optical and infrared 
photometric sequences in low-extinction young clusters like the Pleiades have been later confirmed 
by proper motions, radial velocity or the presence of lithium (Zapatero Osorio et al. \cite{osorio97b}; 
Moraux, Bouvier \& Stauffer \cite{moraux01}). Hence, to confirm that our candidates selected by 
means of optical diagrams are not reddened field stars, we collected the point near-infrared observations 
described in Sect. 2.2. 

It can be seen from the $I$ vs. $I-J$ colour-magnitude diagrams in Figure \ref{fig3} that 49 objects show redder colours and magnitudes brighter than the 10 Myr isochrone, which corresponds roughly to the lower envelope of the 
photometric sequence of previously confirmed members; two candidates with $I$$\sim$17.5 and $I-J$$\sim$1.8 lie very 
close to the 10-Myr isochrone in the upper panel of the figure, one of them is the strongly variable object 
S\,Ori J053825.4-024241, which shows a redder $I-J$ colour according to 2MASS photometry; the other candidate lies on the 10-Myr isochrone when using the 
2MASS photometry (bottom panel). All the 51 objects are considered as
very likely members. Only two candidates appear to have colours clearly bluer than the 10 Myr isochrone, and are hereafter considered to be 
probable non-cluster members. 
There are five candidates in common with the BZOR survey; their cluster membership is supported in both studies.
In BZOR we obtained spectroscopy for ten candidates, of which nine were confirmed as cluster members. 
These nine members follow the infrared photometric sequence of the cluster 
(BMZO), while the rejected candidate (S\,Ori 44) has bluer infrared colours. 
In Table \ref{tab2} we list spectral types of 13 objects in the present survey. Spectroscopic data have been obtained from 
B\'ejar (\cite{bejar00}), Zapatero Osorio et al. (\cite{osorio02}) 
and Barrado y Navascu\'es et al. (\cite{barrado03}). These spectra show the presence of 
H$\alpha$ emission in all cases and Li absorption in eight of them observed with higher 
resolution. As a result, all the objects confirmed by the spectroscopy, have also been previously confirmed 
by infrared photometry, which argues in favour of the reliability of our selection criteria. 

 For those objects with available spectroscopy, we have estimated their $R-I$ colour excesses ($E(R-I)$), 
 according to the $R-I$ colour expected for their spectral types, using relations derived by 
 Kirkpatrick \& McCarthy (\cite{kirk94}), and the $R-I$ photometry obtained here.
 From Table \ref{tab2}, we can see that all the 13 objects 
show a visual extinction lower than 1 ($E(R-I)<0.27$, $A_{V}<1$) and all except one 
show an extinction lower than 0.25 $(E(R-I)<0.07$, $A_{V}<0.25$).  
We have obtained $K_{\rm s}$ photometry for several objects in both surveys. Figure \ref{fig4} shows the $I$ vs. $I$-$K_{\rm s}$ diagram
of present paper candidates. These data confirmed our previous results in the $J$ band. 
In conclusion, although we cannot say, for individual sources, that each candidate belongs to $\sigma$ Orionis 
until we have confirmed their cool temperature and youth spectroscopically, or measured their proper motions, 
we are confident that most of the candidates confirmed with infrared data are bona fide cluster  
members. 

\subsection{Near-Infrared Excesses and the Possible Presence of Discs}

\begin{figure}
\resizebox{\hsize}{!}
{\includegraphics[]{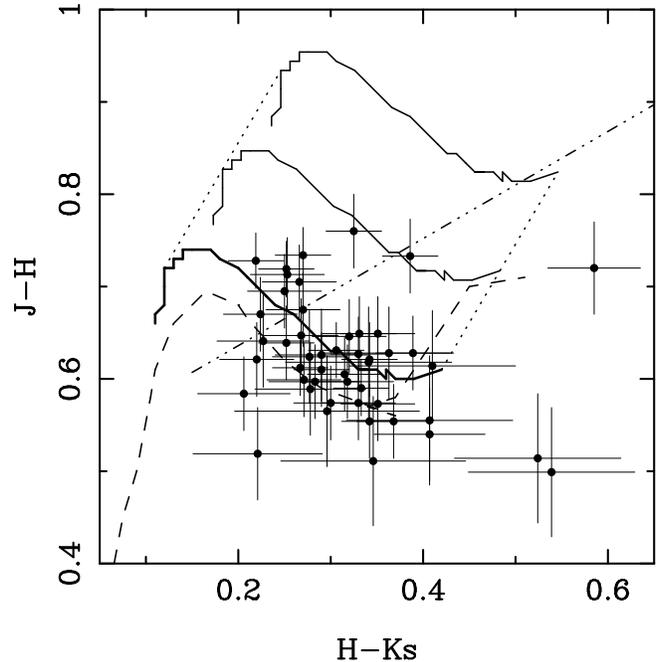}}
\caption{$J-H$ vs.\ $H-K_{\rm s}$ colour--colour diagram of selected candidates with available
$JHK_{\rm s}$ photometry  with accuracy better than 0.1 mag. from 2MASS. 
The 3 Myr Next Gen isochrone from the Lyon group (Baraffe et al. \cite{baraffe98}), reddened   
by visual extinctions of $A_{V}$ = 0, 1and 2, are 
plotted in solid lines from bottom to top. The field dwarf sequence (dashed line) from Bessell \& Brett (\cite{bessell88}) and 
Kirkpatrick \& McCarthy (\cite{kirk94}) and the CTT star loci (dash-dotted line) from Meyer et al. (\cite{meyer97}) 
are also indicated.
}
\label{fig5}
\end{figure}

Using the available $JHK_{\rm s}$ photometry in the 2MASS catalogue, 
we have constructed the $H-K_{\rm s}$ vs.\ $J-K_{\rm s}$ 
colour-colour diagram shown in Figure \ref{fig5}, where we also present Next Gen models (solid lines) 
from the Lyon group (Baraffe et al. \cite{baraffe98}) displaced according to different extinctions, the field dwarf 
sequence (dashed line) from Bessell \& Brett (\cite{bessell88}) and Kirkpatrick \& McCarthy (\cite{kirk94}) and the Classical T Tauri (CTT) 
star loci (dash-dotted line) from Meyer et al. (\cite{meyer97}). Only photometry more accuracy than 0.1 mag. is
plotted in Figure \ref{fig5}. From this colour--colour diagram we can see that 
three out of 48 of our candidates show a near-infrared excess (S\,Ori J054001.9-022133,
S\,Ori J053943.2-023243 and 
S\,Ori J053825.4-024241). The first of these is an M4 low-mass star with an $R-I$ colour of 1.52 $\pm$ 0.06, consistent with its spectral 
type and with the presence of negligible interstellar extinction (see Table \ref{tab2}). 
This suggests that the infrared excess is caused by the presence of a disc. For the second, spectroscopy is unavailable 
and we can not estimate their extinction, so we do not know if the infrared excess is caused by the presence of a disc, interstellar extinction or a 
local small cloud within the cluster. 
The last is the strongly variable brown dwarf candidate with an $R-I$ colour of 1.80$\pm$0.08 and no available 
spectroscopy. Its strong infrared excess can not be explained by the existence of normal interstellar 
reddening and it is most probably related to the presence of a disc. The fraction of objects found with 
near-infrared excesses is in good agreement with previous studies by Oliveira et al. (\cite{oliveira02}), 
who found excesses in only two out of 34 cluster members and Barrado y Navascu\'es et al. (\cite{barrado03}), 
who found excesses in 5--9\% of cluster members.

\section{Conclusions} 

In this paper we present an $RI$ survey of the $\sigma$ Orionis cluster, covering an area of 
430 arcmin$^2$. We have selected 53 candidates in the magnitude range $14<I<20$ that follow the optical 
photometric sequence of the cluster corresponding to masses in the range 0.35--0.015 $M_{\odot}$. 
All but two of the candidates follow the cluster sequence in the infrared and are considered to be likely cluster members: 
around 31 are stars and 20 are brown dwarfs. 
The available spectroscopy for some of these objects confirms them as bona fide members. Using near-infrared photometry from 2MASS, 
we conclude that only three of the 48 (6 \%) candidates show near-infrared excesses possibly related 
to the presence of discs.   

%Our studies in the spatial distribution of cluster members shows that the density of objects is decreasing 
%with the distance to the center. If we represent this fall by a linear function it can be seen that 
%the number of objects in both surveys represent around 40 \% of the total number of cluster members 
%in this range of magnitudes, which argue in favor that both surveys cover a representative part 
%of the cluster. 

%The mass spectrum behaviour of the survey presented in this paper is a growing function to lower mass 
%objects until  the completeness mass 0.030 M$_{\odot}$. This mass spectrum can be represented in the range from 
%0.25 M$_{\odot}$ to 0.030 M$_{\odot}$ by a potential law of exponent around 0.5$\pm$0.2. 
%Comparison with other works in the Pleiades, $\rho$ Ophiucho and 
%IC 348 are in good agreement with our results, indicating that substellar objects can be very numerous 
%over all the Galaxy. 

\acknowledgements
We thank J. Licandro for his help in the acquisition of infrared data at the 
Carlos S\'anchez Telescope and J. A. Caballero and the anonymous referee for their useful comments. We thank I. Baraffe and the Lyon group, 
F. D'Antona and A. Burrows for sending us electronic versions of their recent models. We are indebted to T. Mahoney 
for the english revision of this manuscript. This 
work is based on observations obtained at the  TCS and IAC-80 telescope operated by 
the Instituto de Astrof\'{\i}sica de Canarias at the Spanish 
Observatorio del Teide (Tenerife, Spain).  Partial financial support was provided 
by the Spanish MCYT project AYA2001-1657. This publication makes use of data products from the Two Micron 
All Sky Survey, which is a joint project of the University of Massachusetts and the Infrared
Processing and Analysis Center/California Institute of Technology, funded by the National Aeronautics and 
Space Administration and the National Science Foundation. This research has made use of the SIMBAD database, 
operated at CDS, Strasbourg, France.

\clearpage

\begin{table}[h]
\caption{ Sigma Ori members candidates studied in this paper}
\label{tab2}
\scriptsize
\begin{tabular}{lcccccccccccccc}\hline
 Name  & Other &Finder chart   & $I$  & $R-I$ & $J$  & $I-J$ & $I-K_{\rm s}$ &  \multicolumn{3}{c}{\underline{\qquad \qquad 2MASS photometry \qquad \qquad }} &   Sp. Type       &  E($R-I$)  & R.A.     & DEC. \\
 IAU   & Name  &id. num.       &      &       &      &       &       & $I-J$ & $J-H$ & $H-K_{\rm s}$&                  &  mag & (J2000)  & (J2000) \\
\hline
S\,Ori J053951.1-024944         && 50 & 14.14$\pm$0.04  &  1.38$\pm$0.06  &  12.79$\pm$0.04  & 1.35$\pm$0.06  &2.24$\pm$0.06	   &	1.34$\pm$0.05  &    0.67$\pm$0.04  &  0.22$\pm$0.04 &&& 5 39 51.1 & -2 49 44\\
S\,Ori J053958.2-022619$^{a}$   && 51 & 14.19$\pm$0.04  &  1.42$\pm$0.06  &  12.76$\pm$0.04  & 1.43$\pm$0.06  &2.31$\pm$0.06	   &	1.41$\pm$0.05  &    0.73$\pm$0.03  &  0.22$\pm$0.03 &  M3.0 &  0.06$\pm$0.06& 5 39 58.2 & -2 26 19\\
S\,Ori J053948.1-024557         && 52 & 14.20$\pm$0.04  &  1.25$\pm$0.06  &  12.85$\pm$0.05  & 1.35$\pm$0.06  &                    &	1.28$\pm$0.05  &    0.64$\pm$0.04  &  0.25$\pm$0.03 &&& 5 39 48.1 & -2 45 57\\
S\,Ori J054001.9-022133$^{a,b}$ && 53 & 14.32$\pm$0.04  &  1.52$\pm$0.06  &  12.44$\pm$0.04  & 1.88$\pm$0.06  &3.08$\pm$0.06       &	1.98$\pm$0.05  &    0.76$\pm$0.04  &  0.32$\pm$0.03 &  M4.0 & -0.07$\pm$0.06& 5 40 01.9 & -2 21 33\\
S\,Ori J053952.3-023615         && 54 & 14.37$\pm$0.04  &  1.32$\pm$0.06  &  12.65$\pm$0.05  & 1.72$\pm$0.06  &                    &	1.48$\pm$0.05  &    0.70$\pm$0.04  &  0.25$\pm$0.04 &&& 5 39 52.3 & -2 36 15\\
S\,Ori J053820.2-023802$^{a,c}$ && 55 & 14.32$\pm$0.06  &  1.69$\pm$0.08  &  12.59$\pm$0.05  & 1.73$\pm$0.08  &                    &	1.74$\pm$0.07  &    0.72$\pm$0.03  &  0.25$\pm$0.03 &  M4.0 &  0.01$\pm$0.06& 5 38 20.2 & -2 38 02\\
S\,Ori J053836.7-024414         && 56 & 14.38$\pm$0.06  &  1.70$\pm$0.08  &  12.55$\pm$0.05  & 1.83$\pm$0.08  &                    &	1.84$\pm$0.07  &    0.65$\pm$0.04  &  0.27$\pm$0.04 &&& 5 38 36.7 & -2 44 14\\
S\,Ori J053827.5-023504$^{a,d}$ && 57 & 14.41$\pm$0.06  &  1.42$\pm$0.08  &  13.12$\pm$0.05  & 1.29$\pm$0.08  &                    &	1.58$\pm$0.07  &    0.71$\pm$0.04  &  0.25$\pm$0.04 &  M3.5 & -0.18$\pm$0.06& 5 38 27.5 & -2 35 04\\
S\,Ori J053820.5-023409$^{a,c}$ && 58 & 14.49$\pm$0.06  &  1.68$\pm$0.08  &  12.93$\pm$0.05  & 1.56$\pm$0.08  &                    &    1.84$\pm$0.07  &    0.73$\pm$0.03  &  0.27$\pm$0.03 &  M4.0 &  0.00$\pm$0.06& 5 38 20.5 & -2 34 09\\
S\,Ori J053951.7-022247$^{a}$   && 59 & 14.59$\pm$0.04  &  1.91$\pm$0.06  &  12.62$\pm$0.04  & 1.97$\pm$0.06  &3.16$\pm$0.07       &	1.99$\pm$0.05  &    0.59$\pm$0.03  &  0.33$\pm$0.03 &  M5.5 & -0.12$\pm$0.06& 5 39 51.7 & -2 22 47\\
S\,Ori J053943.2-023243         && 60 & 14.70$\pm$0.06  &  1.64$\pm$0.08  &  13.04$\pm$0.05  & 1.66$\pm$0.08  &                    &    1.67$\pm$0.07  &    0.73$\pm$0.04  &  0.39$\pm$0.03 &&& 5 39 43.2 & -2 32 43\\
S\,Ori J053946.6-022631         && 61 & 14.83$\pm$0.04  &  1.86$\pm$0.06  &  12.91$\pm$0.05  & 1.92$\pm$0.06  &                    &    1.84$\pm$0.05  &    0.63$\pm$0.03  &  0.30$\pm$0.03 &&& 5 39 46.6 & -2 26 31\\
S\,Ori J053817.8-024050         && 62 & 14.87$\pm$0.06  &  1.86$\pm$0.08  &  13.04$\pm$0.05  & 1.83$\pm$0.08  &                    &    1.67$\pm$0.07  &    0.62$\pm$0.04  &  0.34$\pm$0.03 &&& 5 38 17.8 & -2 40 50\\
S\,Ori J054009.3-022507         && 63 & 15.00$\pm$0.04  &  1.75$\pm$0.06  &  13.13$\pm$0.07  & 1.87$\pm$0.08  &2.94$\pm$0.04       &	1.85$\pm$0.05  &    0.65$\pm$0.04  &  0.35$\pm$0.04 &&& 5 40 09.3 & -2 25 07\\
S\,Ori J053823.6-024132         && 64 & 14.93$\pm$0.06  &  1.81$\pm$0.08  &  13.31$\pm$0.05  & 1.62$\pm$0.08  &                    &    1.64$\pm$0.07  &    0.55$\pm$0.04  &  0.34$\pm$0.03 &&& 5 38 23.6 & -2 41 32\\
S\,Ori J053715.2-024202$^{a}$   && 65 & 15.07$\pm$0.04  &  1.62$\pm$0.06  &  13.59$\pm$0.05  & 1.48$\pm$0.06  &2.49$\pm$0.03       &	1.45$\pm$0.05  &    0.62$\pm$0.04  &  0.22$\pm$0.04 &  M4.0 &  0.03$\pm$0.06& 5 37 15.2 & -2 42 02\\
S\,Ori J053957.5-023212         && 66 & 15.02$\pm$0.06  &  1.80$\pm$0.08  &  13.20$\pm$0.05  & 1.82$\pm$0.08  &                    &    1.71$\pm$0.07  &    0.63$\pm$0.04  &  0.33$\pm$0.03 &&& 5 39 57.5 & -2 32 12\\
S\,Ori J053949.4-022346$^{a}$   && 67 & 15.14$\pm$0.04  &  1.80$\pm$0.06  &  13.37$\pm$0.04  & 1.77$\pm$0.06  &2.99$\pm$0.07       &	1.74$\pm$0.05  &    0.65$\pm$0.04  &  0.32$\pm$0.04 &  M4.0 &  0.21$\pm$0.06& 5 39 49.4 & -2 23 46\\
S\,Ori J053823.3-024414         && 68 & 15.05$\pm$0.06  &  1.70$\pm$0.08  &  13.35$\pm$0.05  & 1.70$\pm$0.08  &                    &    1.59$\pm$0.07  &    0.61$\pm$0.04  &  0.29$\pm$0.03 &&& 5 38 23.3 & -2 44 14\\
S\,Ori J053954.2-022733         && 69 & 15.15$\pm$0.04  &  1.75$\pm$0.06  &  13.51$\pm$0.04  & 1.64$\pm$0.06  &2.46$\pm$0.05       &	1.69$\pm$0.05  &    0.58$\pm$0.04  &  0.21$\pm$0.05 &&& 5 39 54.2 & -2 27 33\\
S\,Ori J054007.1-023245         && 70 & 15.10$\pm$0.06  &  1.72$\pm$0.08  &  13.45$\pm$0.04  & 1.65$\pm$0.07  &2.63$\pm$0.08       &	1.68$\pm$0.07  &    0.61$\pm$0.03  &  0.27$\pm$0.03 &&& 5 40 07.1 & -2 32 45\\
S\,Ori J053956.4-023804         && 71 & 15.21$\pm$0.04  &  1.71$\pm$0.06  &  13.32$\pm$0.06  & 1.89$\pm$0.07  &                    &    1.86$\pm$0.05  &    0.55$\pm$0.04  &  0.37$\pm$0.04 &&& 5 39 56.4 & -2 38 04\\
S\,Ori J053950.6-023414         && 72 & 15.37$\pm$0.06  &  1.78$\pm$0.08  &  13.77$\pm$0.05  & 1.60$\pm$0.08  &                    &    1.69$\pm$0.07  &    0.67$\pm$0.04  &  0.27$\pm$0.04 &&& 5 39 50.6 & -2 34 14\\
S\,Ori J053845.9-024523         && 73 & 15.52$\pm$0.06  &  1.84$\pm$0.08  &  13.46$\pm$0.05  & 2.06$\pm$0.08  &                    &    1.96$\pm$0.07  &    0.60$\pm$0.04  &  0.27$\pm$0.04 &&& 5 38 45.9 & -2 45 23\\
S\,Ori J053811.9-024557         && 74 & 15.72$\pm$0.06  &  1.87$\pm$0.08  &  13.68$\pm$0.06  & 2.04$\pm$0.08  &                    &    1.80$\pm$0.07  &    0.63$\pm$0.05  &  0.29$\pm$0.05 &&& 5 38 11.9 & -2 45 57\\
S\,Ori J054005.3-023052$^{a}$   && 75 & 15.81$\pm$0.06  &  1.89$\pm$0.08  &  13.93$\pm$0.04  & 1.88$\pm$0.07  &2.89$\pm$0.10       &	1.86$\pm$0.07  &    0.57$\pm$0.04  &  0.30$\pm$0.04 &  M5.0 & -0.09$\pm$0.06& 5 40 05.3 & -2 30 52\\
S\,Ori J053850.6-024244         && 76 & 15.84$\pm$0.06  &  1.85$\pm$0.08  &  13.75$\pm$0.05  & 2.09$\pm$0.08  &                    &    2.00$\pm$0.07  &    0.60$\pm$0.04  &  0.28$\pm$0.05 &&& 5 38 50.6 & -2 42 44\\
S\,Ori J053833.9-024508         && 77 & 15.98$\pm$0.06  &  1.89$\pm$0.07  &  14.28$\pm$0.06  & 1.70$\pm$0.08  &                    &    1.73$\pm$0.07  &    0.57$\pm$0.04  &  0.33$\pm$0.04 &&& 5 38 33.9 & -2 45 08\\
S\,Ori J053826.8-023846         && 78 & 16.12$\pm$0.06  &  1.90$\pm$0.08  &  14.20$\pm$0.05  & 1.92$\pm$0.08  &                    &    2.01$\pm$0.07  &    0.62$\pm$0.05  &  0.28$\pm$0.04 &&& 5 38 26.8 & -2 38 46\\
S\,Ori J053848.1-024401         && 79 & 16.13$\pm$0.06  &  1.91$\pm$0.08  &  14.02$\pm$0.05  & 2.11$\pm$0.08  &                    &    2.06$\pm$0.07  &    0.60$\pm$0.04  &  0.31$\pm$0.04 &&& 5 38 48.1 & -2 44 01\\
S\,Ori J053944.4-022445$^{*}$   & S\,Ori 10 & 10 &  16.25$\pm$0.04  &  1.88$\pm$0.06  &  14.12$\pm$0.02  & 2.13$\pm$0.04  &        &	2.08$\pm$0.05  &    0.63$\pm$0.04  &  0.39$\pm$0.04 &&& 5 39 44.4 & -2 24 43\\
S\,Ori J053944.2-023305$^{*,e}$ & S\,Ori 11 & 11 &  16.39$\pm$0.06  &  2.04$\pm$0.08  &  14.47$\pm$0.05  & 1.92$\pm$0.08  &        &	2.10$\pm$0.07  &    0.57$\pm$0.04  &  0.35$\pm$0.04 &  M6.0 & -0.22$\pm$0.06& 5 39 44	& -2 33 03\\
S\,Ori J053838.6-024157         && 80 & 16.52$\pm$0.06  &  1.82$\pm$0.08  &  14.54$\pm$0.05  & 1.98$\pm$0.08  &	                   &	1.96$\pm$0.07  &    0.60$\pm$0.04  &  0.31$\pm$0.05 &&& 5 38 38.6 & -2 41 57\\
S\,Ori J053826.2-024041$^{e}$   && 81 & 16.87$\pm$0.06  &  2.13$\pm$0.08  &  14.94$\pm$0.06  & 1.93$\pm$0.08  &	                   &    1.96$\pm$0.07  &    0.63$\pm$0.05  &  0.36$\pm$0.07 &  M8.0 &	  & 5 38 26.2 & -2 40 41\\
S\,Ori J053954.3-023720         && 82 & 17.03$\pm$0.04  &  2.10$\pm$0.06  &  14.67$\pm$0.05  & 2.36$\pm$0.06  &	                   &    2.28$\pm$0.05  &    0.54$\pm$0.05  &  0.41$\pm$0.06 &&& 5 39 54.3 & -2 37 20\\
S\,Ori J053817.3-024024$^{*,a}$ & S\,Ori 27 & 27 &  17.00$\pm$0.06  &  2.16$\pm$0.08  &  14.82$\pm$0.05  & 2.18$\pm$0.08  &        &	2.17$\pm$0.07  &    0.52$\pm$0.05  &  0.22$\pm$0.07 &  M6.5 & -0.30$\pm$0.06& 5 38 17.3 & -2 40 24\\
S\,Ori J053820.8-024613$^{*,e}$ & S\,Ori 31 & 31 &  17.20$\pm$0.06  &  2.35$\pm$0.08  &  15.11$\pm$0.05  & 2.09$\pm$0.08  &        &	2.01$\pm$0.07  &    0.61$\pm$0.06  &  0.41$\pm$0.09 &  M7.0 & -0.21$\pm$0.06& 5 38 20.8 & -2 46 13\\
S\,Ori J053844.4-024037         && 83 & 17.28$\pm$0.06  &  2.32$\pm$0.08  &  14.92$\pm$0.20  & 2.36$\pm$0.21  &                    &    2.48$\pm$0.07  &    0.59$\pm$0.05  &  0.28$\pm$0.06 &&& 5 38 44.4 & -2 40 37\\
S\,Ori J053943.5-024731$^{*}$   & S\,Ori 32 & 32 &  17.57$\pm$0.04  &  2.12$\pm$0.06     &   15.16$\pm$0.15	& 2.41$\pm$0.16&   &	2.23$\pm$0.06  &    0.55$\pm$0.07  &  0.41$\pm$0.09 &&& 5 39 43.5 & -2 47 32 \\
S\,Ori J053700.9-023856         && 84 & 17.41$\pm$0.04 &   2.03$\pm$0.06  &  15.51$\pm$0.15  & 1.90$\pm$0.15  &&    1.95$\pm$0.08  &    0.54$\pm$0.11  &  0.29$\pm$0.15  &&& 5 37 00.9 & -2 38 56\\
S\,Ori J053821.3-023336$^{**}$  && 85 & 17.58$\pm$0.06  &  2.13$\pm$0.08  &  15.40$\pm$0.07  & 2.18$\pm$0.09  &                    &	2.22$\pm$0.07  &    0.56$\pm$0.06  &  0.30$\pm$0.10 &&& 5 38 21.4 & -2 33 36\\
S\,Ori J053805.5-023557         && 86 & 17.66$\pm$0.06  &  2.10$\pm$0.11  &  15.11$\pm$0.08  & 2.55$\pm$0.10  & 	           &	2.38$\pm$0.07  &    0.51$\pm$0.07  &  0.52$\pm$0.09 &&& 5 38 05.5 & -2 35 57\\
S\,Ori J054004.5-023642         && 87 & 17.77$\pm$0.04  &  2.16$\pm$0.06  &  15.40$\pm$0.06  & 2.37$\pm$0.07  &                    &	2.46$\pm$0.06  &    0.50$\pm$0.07  &  0.54$\pm$0.09 &&& 5 40 04.5 & -2 36 42\\
S\,Ori J053853.8-024459         && 88 & 17.78$\pm$0.06  &  2.08$\pm$0.08  &  15.47$\pm$0.05  & 2.31$\pm$0.08  &                    &	2.33$\pm$0.07  &    0.51$\pm$0.07  &  0.35$\pm$0.10 &&& 5 38 53.8 & -2 44 59\\
S\,Ori J053818.2-023539         && 89 & 17.96$\pm$0.06  &  2.23$\pm$0.08  &  15.41$\pm$0.05  & 2.55$\pm$0.08  &                    &	2.51$\pm$0.07  &    0.62$\pm$0.06  &  0.34$\pm$0.09 &&& 5 38 18.2 & -2 35 39\\
S\,Ori J053948.1-022914$^{**,e}$&& 90 & 18.67$\pm$0.07  &  2.33$\pm$0.10  &  16.16$\pm$0.03  & 2.51$\pm$0.08  &                    &    2.25$\pm$0.12  &    0.83$\pm$0.13  &  0.40$\pm$0.17 &  M7.0 & -0.23$\pm$0.09& 5 39 48.1 & -2 29 14\\
S\,Ori J053710.0-024302$^{f}$   && 92 & 19.95$\pm$0.09  &  2.94$\pm$0.33  &  16.77$\pm$0.30  & 3.18$\pm$0.30  &4.6$\pm$0.4         &		       &		   & 	            &  M8.0 &     & 5 37 10.0 & -2 43 02\\
S\,Ori J053834.5-024109         && 94 &	14.83$\pm$0.06  &  1.41$\pm$0.08  &  13.05$\pm$0.05  & 1.78$\pm$0.08  &                    &    1.73$\pm$0.07  &    0.65$\pm$0.04  &  0.33$\pm$0.04 &&& 5 38 34.5 & -2 41 09\\
S\,Ori J053844.4-024030         && 95 &	15.02$\pm$0.06  &  1.49$\pm$0.08  &  13.34$\pm$0.07  & 1.68$\pm$0.09  &                    & 	1.65$\pm$0.07  &    0.64$\pm$0.05  &  0.23$\pm$0.05 &&& 5 38 44.4 & -2 40 30\\
S\,Ori J053816.0-023805         && 96 &	15.19$\pm$0.06  &  1.57$\pm$0.08  &                  &                &                    & 	1.61$\pm$0.07  &    0.70$\pm$0.04  &  0.27$\pm$0.04 &&& 5 38 16.0 & -2 38 05\\
S\,Ori J053825.4-024241         && 97 &	16.86$\pm$0.06  &  1.80$\pm$0.08  &  15.09$\pm$0.11  & 1.77$\pm$0.13  &                    & 	1.98$\pm$0.07  &    0.72$\pm$0.05  &  0.58$\pm$0.05 &&& 5 38 25.4 & -2 42 41\\

\hline

&& & & Probably & non & members & & && &\\
\hline
S\,Ori J054004.9-024656         && 91 & 19.91$\pm$0.07 &   2.64$\pm$0.28  &  17.38$\pm$0.20  & 2.53$\pm$0.20&&&&&&& 5 40 04.9  &-2 46 56\\ 
S\,Ori J053959.5-022146         && 93 & 19.97$\pm$0.07 &   2.39$\pm$0.25  &  17.27$\pm$0.35  & 2.70$\pm$0.35  &3.6$\pm$0.5&&&&&& 5 39 59.5 & -2 21 46\\

\hline\hline
\end{tabular}

* IAU name from BZOR

** IAU name from BMZO

a. Spectroscopic data from Zapatero Osorio et al. \cite{osorio02}. All these objects have Li.

b. Haro 5-36, from Haro, M. \cite{haro53}

c. X-ray source from Wolk \cite{wolk96}

d. Kiso A-0976 329 from Wiramihardja et al. \cite{wira91}

e. Spectroscopic data from Barrado y Navascu\'es et al. \cite{barrado03}

f. Spectroscopic data from B\'ejar  \cite{bejar00}
\end{table}

\clearpage

\begin{appendix}

\section{Finder charts.}

\begin{figure*}
\resizebox{\hsize}{!}
{\includegraphics[]{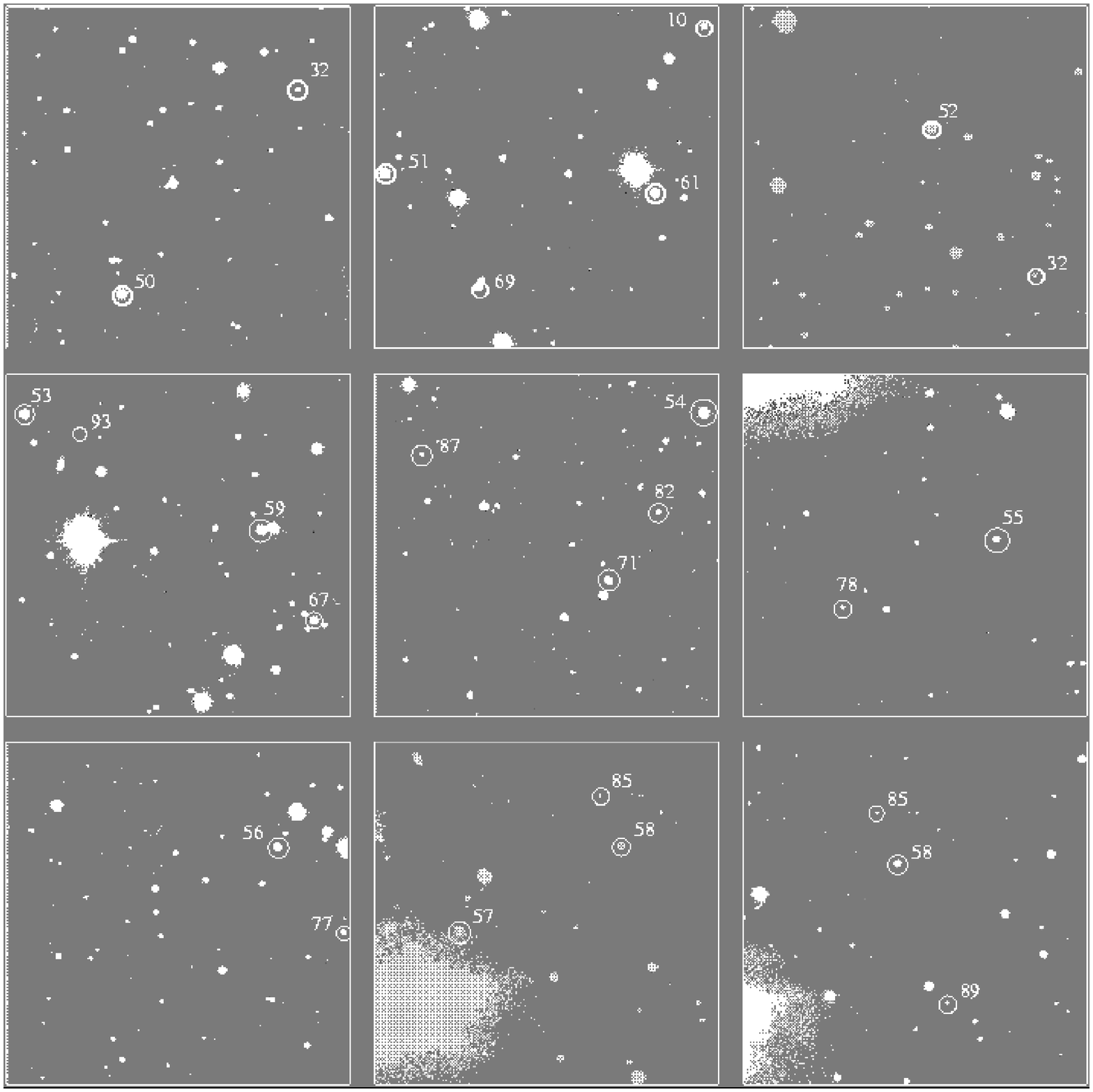}}
\caption{Finder charts in the 
$I$-band (3.7\arcmin$\times$3.7\arcmin). North is up and East is 
left. 
}
\label{figA1}
\end{figure*}

\begin{figure*}
\resizebox{\hsize}{!}
{\includegraphics[]{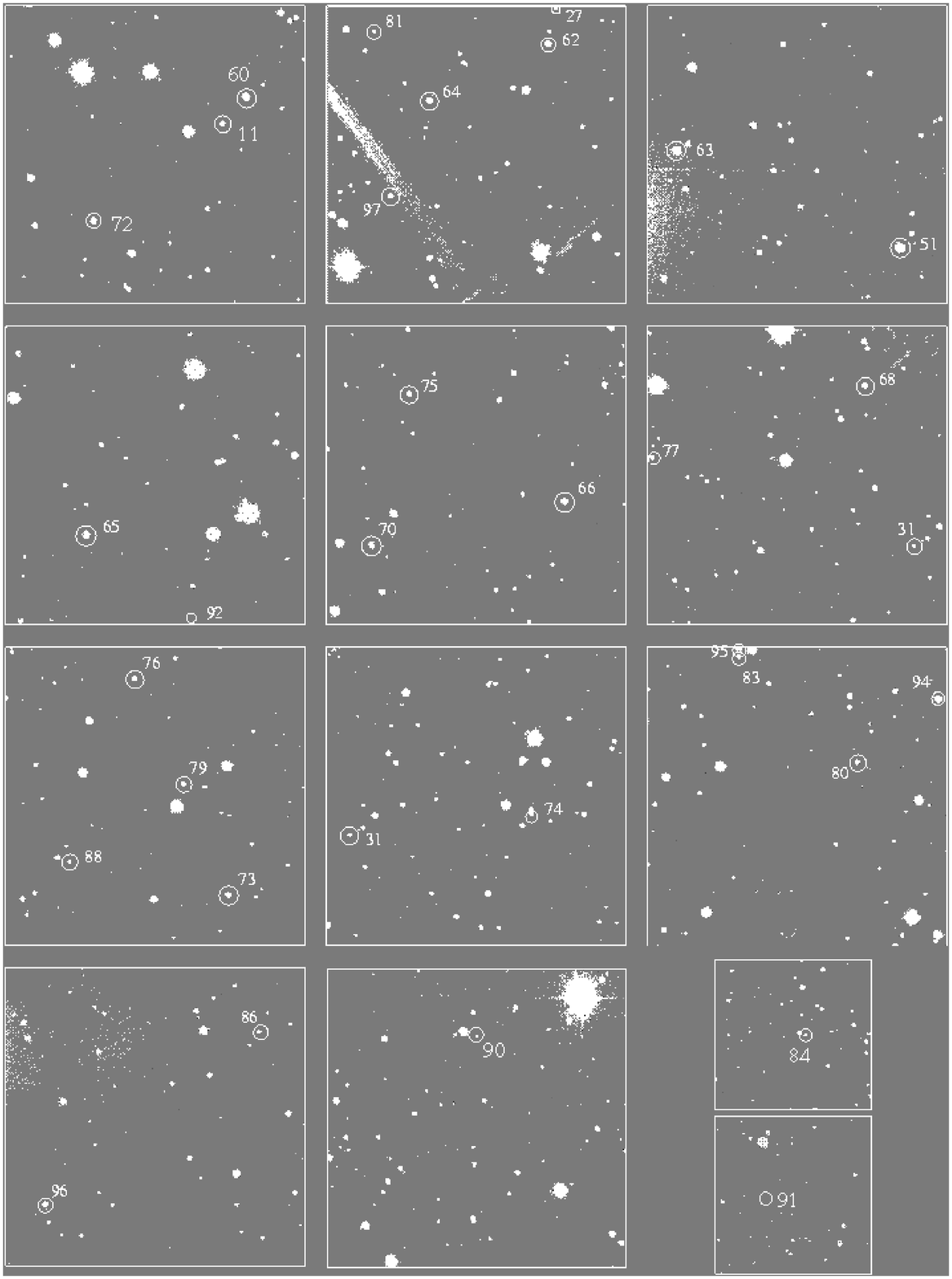}}
\caption{Finder charts in the 
$I$-band (3.7\arcmin$\times$3.7\arcmin). North is up and East is 
left (continuation). 
}
\label{figA2}
\end{figure*}

\end{appendix}

\end{document}